\documentclass{ws-ijmpcs}

\begin{document}

\markboth{V. I. Mokeev and I. G. Aznauryan}
{Studies of N$^*$ structure from the CLAS meson electroproduction data}

%
\catchline{}{}{}{}{}
%

\title{Studies of N$^*$ structure from the CLAS meson electroproduction data 
}

\author{V. I. Mokeev
}

\address{Thomas Jefferson National Accelerator Facility,\\
Newport News, VA, 23606,
USA
\\
mokeev@jlab.org}

\author{I. G. Aznauryan}

\address{Yerevan Physics Institute,\\
Yerevan, 375036, Armenia\\}



\maketitle

\begin{history}
\received{Day Month Year}
\revised{Day Month Year}
\end{history}

\begin{abstract}
The transition $\gamma_{v}pN^*$ amplitudes (electrocouplings) for prominent excited nucleon states obtained  in a wide area of photon virtualities  offer valuable information for the exploration of the N$^*$ structure at different distances and allow us to access the complex dynamics of non-perturbative strong interaction. The current status in the studies of $\gamma_{v}pN^*$  electrocouplings  from the data on exclusive meson electroproduction off protons measured with the CLAS detector at Jefferson Lab is presented. The impact of these results on exploration of the N$^*$ structure is discussed. 

\end{abstract}

\ccode{PACS numbers: 11.55.Fv, 13.40.Gp, 13.60.Le, 14.20.Gk}

\section{Introduction}

Studies of $\gamma_{v}pN^*$ transition amplitudes (electrocouplings) represent an important part of the efforts in  exploration of 
strong interaction dynamics in the non-perturbative regime of strong quark-gluon coupling. The  evolution of $\gamma_{v}pN^*$ electrocouplings 
with photon virtualities Q$^2$ opens up access to the relevant degrees of freedom in the N$^*$ structure at different distance scales. 
It allows us to explore how the strong interaction of quarks and gluons generates different $N^*$ states and how it evolves with distance from the pQCD 
regime to quark-gluon confinement \cite{CLAS12,CL12}. In this contribution we discuss the current status of the $\gamma_{v}pN^*$ electrocoupling studies from exclusive meson electroproduction off protons and their impact on our understanding of the N$^*$ structure.

\section{Studies of $\gamma_{v}pN^*$ electrocouplings in exclusive meson electroproduction}

Nucleon resonance electroexcitations are described by two transverse ($A_{1/2}(Q^2)$, $A_{3/2}(Q^2)$) and one longitudinal ($S_{1/2}(Q^2)$) 
electrocoupling amplitudes \cite{azbu12}. These electrocouplings offer an access to the resonance structure. The CLAS detector afforded excellent opportunities to study N$^*$ electroexcitation with great precision and has contributed the lion's share of the world's data on all essential exclusive meson electroproduction channels in the resonance excitation region. CLAS provides nearly complete coverage of the final hadron phase space \cite{jlab11}. Electrocouplings have been obtained from the CLAS data of $\pi^+n$, $\pi^0p$ exclusive channels at Q$^2$ up to 5 GeV$^2$, $\eta p$ channel at Q$^2$ $<$ 4.0 GeV$^2$, and from $\pi^+\pi^-p$ electroproduction at Q$^2$ $<$ 1.5 GeV$^2$ \cite{azbu12,jlab11}. Several phenomenological reaction models were developed for evaluation the $\gamma_{v}pN^*$ electrocouplings in independent analyses of N$\pi$ \cite{Az03,Az09,Tia11,Ar06}, N$\eta$  \cite{Az031,De07} and $\pi^{+}\pi^{-}p$ electroproduction \cite{Mo09,Mo12}. Coupled-channel analyses are also making progress toward the extraction of resonance electrocouplings in global multi-channel data fits \cite{CLAS12,Lee13}. 

The $N\pi$ and $\pi^+\pi^-p$ exclusive channels provided a major part of the information on electrocouplings of nucleon resonances.
The CLAS data considerably extends information available on $\pi^{+}n$, $\pi^0p$ electroproduction off protons. A total
of nearly 120,000 data points on unpolarized differential cross sections, longitudinally polarized beam asymmetries, and
longitudinal target and beam-target asymmetries were obtained with the CLAS detector  at W $<$ 1.7 GeV and 0.2 GeV$^2$ $<$ Q$^2$ $<$ 5.0 GeV$^2$ \cite{Az09}. Recently, preliminary $\pi^+n$ electroproduction data (~36,000 data points) have become available at 1.6 GeV $<$ W $<$ 2.0 GeV and 1.8 GeV$^2$ $<$ Q$^2$ $<$ 4.0 GeV$^2$ \cite{park13}. The data were analyzed within the framework of two conceptually
different approaches: a) the unitary isobar model, and b) a model, employing dispersion relations
\cite{Az03}. All well established $N^*$ states in the mass range $M_{N^*}$ $<$ 1.8 GeV were 
incorporated into the $N\pi$ channel analyses. The two approaches provide good  description of the $N\pi$ data in the range: $W$ $<$ 1.7 GeV and $Q^2$ $<$ 5.0 GeV$^2$, resulting in $\chi^2$/d.p. $<$ 2.0 \cite{Az09}. 
This good description 
of a large body of different observables allowed us to obtain reliable information on the $A_{1/2}$, $A_{3/2}$, and $S_{1/2}$ resonance 
electrocouplings. 

The $\pi^+\pi^- p$ electroproduction data from CLAS \cite{Ri03,Fe09} provide for the first time information on 
nine independent one-fold-differential and fully-integrated cross sections in each bin of $W$ and $Q^2$ in a mass range 
$W$ $<$ 2.0 GeV, and at
photon virtualities of 0.25 GeV$^2$ $<$ $Q^2$ $<$ 1.5 GeV$^2$. The analysis of these data allowed us to establish all essential mechanisms contributing to $\pi^+\pi^-p$ electroproduction: $\pi^-\Delta^{++}$, $\pi^+\Delta^0$,
$\rho^0 p$, $\pi^+N(1520)\frac{3}{2}^-$, $\pi^+N(1685)\frac{5}{2}^+$, $\pi^-\Delta(1620)\frac{3}{2}^+$ meson-baryon channels and direct production of the $\pi^+\pi^-p$ final state without formation of intermediate unstable hadrons. The data-driven reaction model JM has been developed \cite{Mo09,Mo12} with the goal of extracting resonance electrocouplings and the $\pi\Delta$ and $\rho p$ hadronic decay widths. The contributions from well established N$^*$ states in the mass range up to 2.0 GeV were included into the amplitudes of $\pi\Delta$ and $\rho p$ meson-baryon channels. Resonant contributions were treated employing unitarized version of the Breit-Wigner ansatz \cite{Mo12}.  
The JM model provides a reasonable description of $\pi^+\pi^- p$ differential cross sections at $W$ $<$ 1.8 GeV and $Q^2$ $<$ 1.5 GeV$^2$. The successful description of  $\pi^+\pi^-p$ electroproduction cross sections
allowed us to isolate the resonant contributions and to determine both resonance  
electrocouplings and $\pi \Delta$ and $\rho p$ decay widths fitting them to the measured observables.

\section{The results on $\gamma_{v}pN^*$ electrocouplings and their impact on the studies of resonance structure }

\begin{figure}[htp]
  \includegraphics[width=4.5cm]{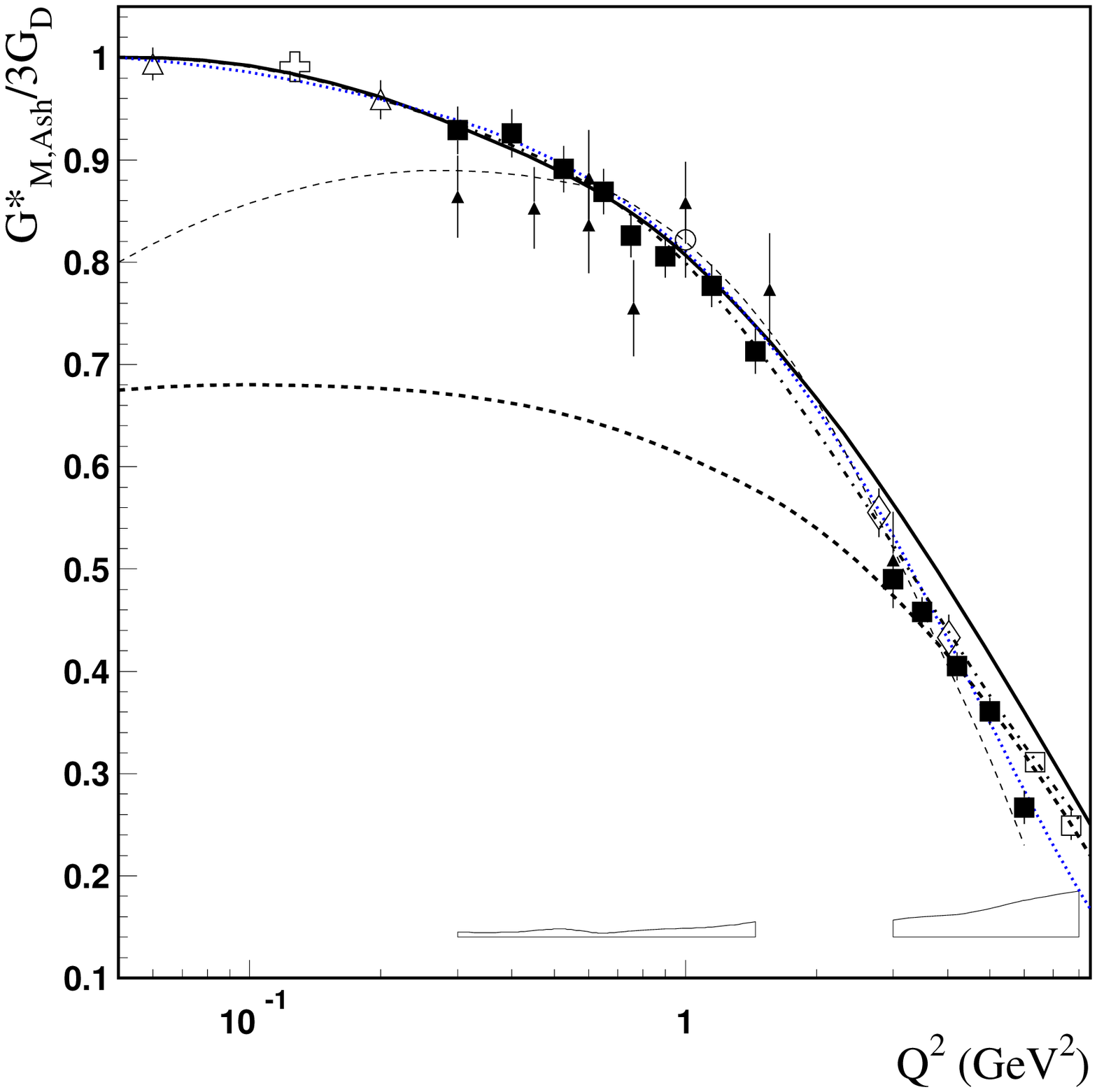}
  \includegraphics[width=6.5cm]{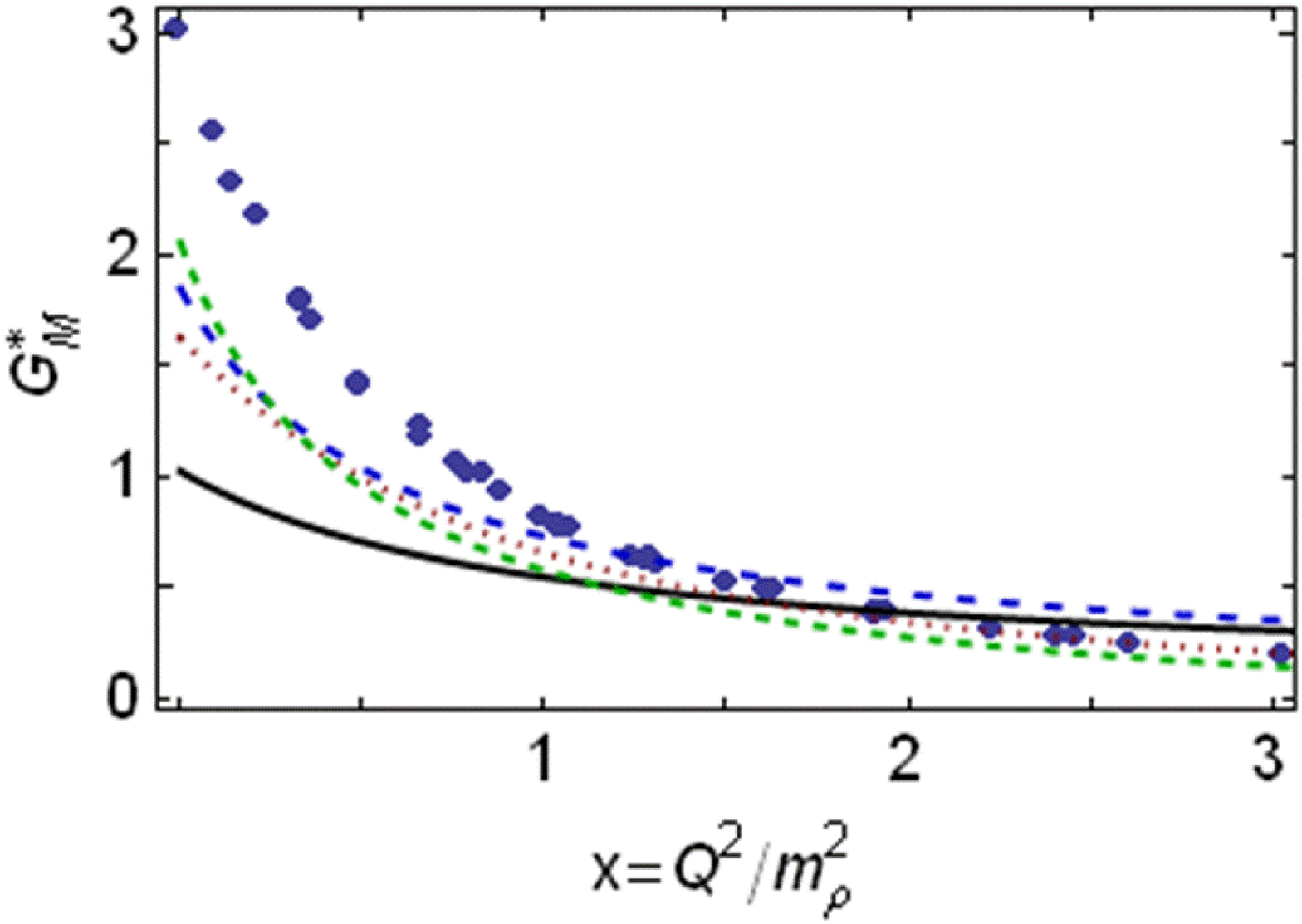}
\vspace*{8pt}
\caption{(Color online) Left: Transition p $\rightarrow$ $\Delta$ magnetic form factor in Ash convention \protect\cite{ash} normalized to the dipole form 3$G_{D}(Q^2)$ . Experimental data are taken from the review \protect\cite{azbu12}.  The dashed and solid curves correspond to quark core and combined quark core and meson baryon cloud contributions \protect\cite{Lee01}. Right: DSEQCD results on quark core contribution to the  p $\rightarrow$ $\Delta$ transition magnetic form factor obtained with contact qq-interaction (solid line) and accounting for a quark anomalous magnetic moment (long-dashed blue line) in comparison with a coupled channel approach \protect\cite{Lee01} inferred from experimental data (short-dashed green line) \protect\cite{cro13}. The DSEQCD expectations for realistic qq-interaction is shown as dotted red line. Experimental values of the dressed magnetic transition form factor are shown as points. \label{p33}}
\end{figure}

The p $\rightarrow$ $\Delta$ magnetic transition form factor in Ash convention \cite{ash} determined from analyses of exclusive $N\pi$
electroproduction data \cite{azbu12} and normalized to the dipole form is shown in Fig.~\ref{p33} (left). Physics analyses of these results revealed the structure of $P_{33}(1232)$ state as combined contribution from external meson-baryon cloud and internal core of three dressed quarks in the ground state bound in the flavor decuplet. The models accounting for only quark degrees of freedom allowed one to describe the data at Q$^2$ $>$ 3.0 GeV$^2$ but underestimate considerably the experimental values at Q$^2$ $<$ 1.0 GeV$^2$.  A good description of the data in the full area of photon virtualities shown in Fig.~\ref{p33} (left) has been achieved accounting for the contribution from both external meson-baryon cloud derived from experimental $N\pi$ data within the coupled-channel approach \cite{Lee01} and the inner quark core.

First evaluations of bare quark core contributions to the p $\rightarrow$ $\Delta$ magnetic transition form factor have recently become available from Dyson-Schwinger Equations of QCD (DSEQCD) \cite{cro13,alk13}. The results \cite{cro13} are shown in 
Fig.~\ref{p33} (right) in comparison with the quark core contribution  employing the coupled-channel approach \cite{Lee01} inferred from the experimental data. Consistent results of both approaches on the quark-core contribution  demonstrate promising potential of DSEQCD in describing the $P_{33}(1232)$ resonance quark content starting from the QCD Lagrangian. DSEQCD models allow us to explore how N$^*$ masses are generated non-perturbatively, relating resonance electrocouplings to dressed quark mass function \cite{CLAS12}.

\begin{figure}[htp]
  \includegraphics[width=5. cm]{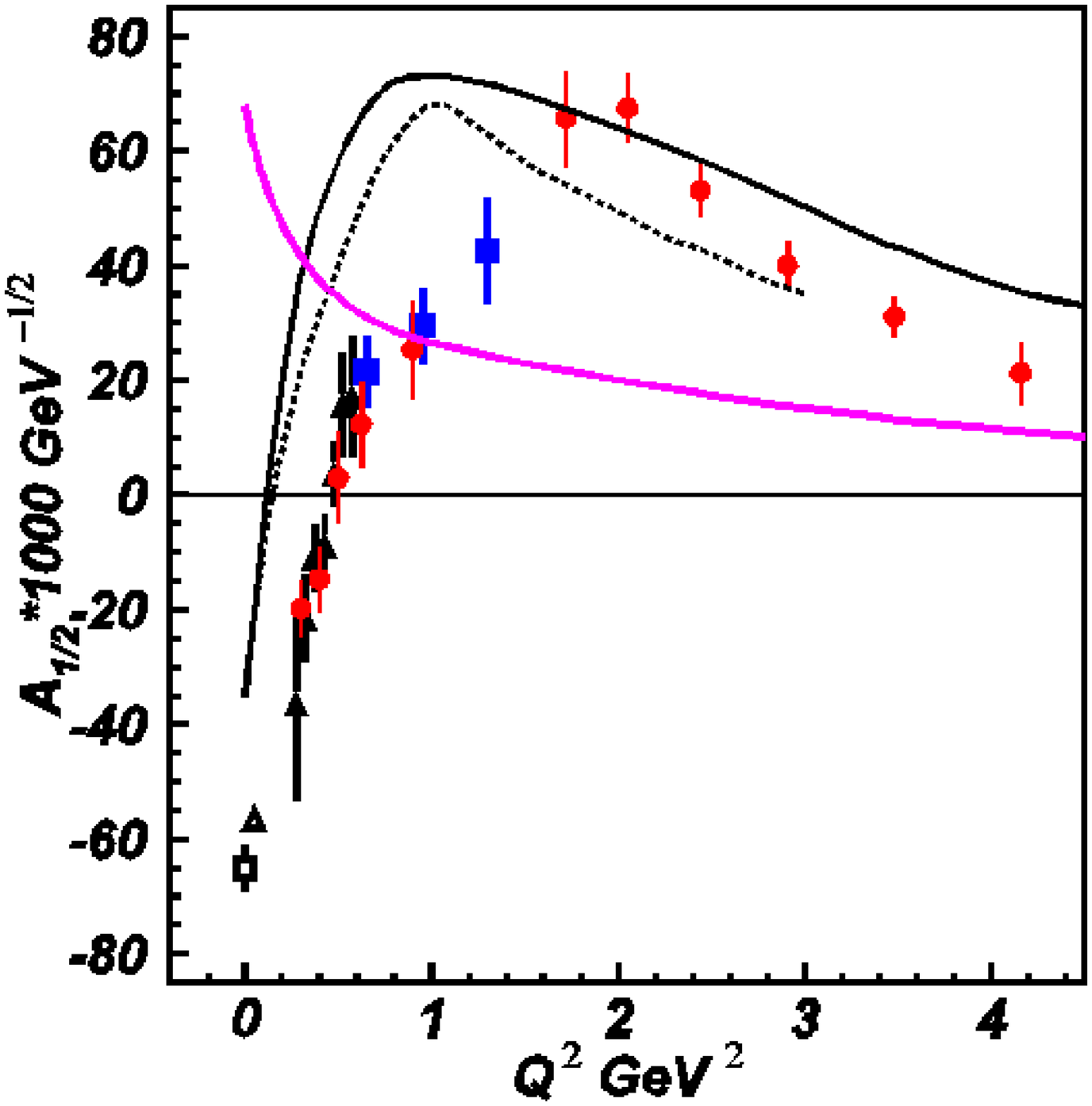}
  \includegraphics[width=5.cm]{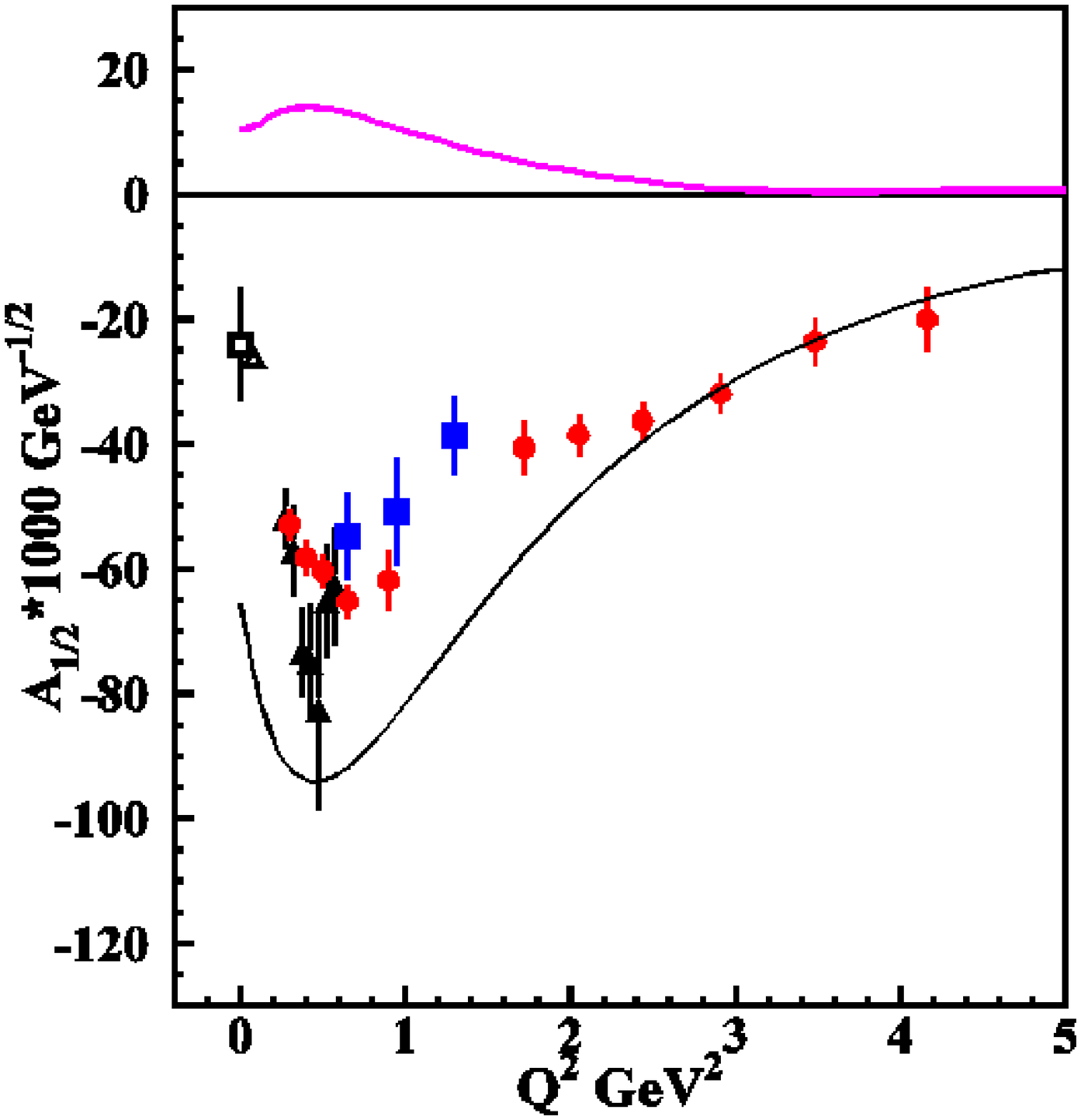}
\vspace*{8pt}
\caption{(Color online) CLAS results on $P_{11}(1440)$ (left) and $D_{13}(1520)$ (right) $A_{1/2}$ electrocouplings obtained from analyses of N$\pi$ electroproduction \protect\cite{Az09} (red circles) and 
$\pi^+\pi^-p$ exclusive channel (black triangles) \protect\cite{Mo12} including recent preliminary results in the area of Q$^2$ from 0.5 to 1.5 GeV$^2$ (blue squares). The results at the photon point are taken from \protect\cite{rpp,Dug09}. Quark core contributions to $P_{11}(1440)$ electrocoupling estimated within the framework of the light-front quark models \protect\cite{Az07,Ca95} are shown in the left part as solid and dashed lines, respectively. The quark core contribution to $D_{13}(1520)$ electrocoupling obtained within the framework of a quark model \protect\cite{Gia12} is shown in the right part by solid line. Absolute values of the meson-baryon cloud contributions inferred from the experimental data in a coupled-channel approach \protect\cite{Lee08} are shown by thick solid lines in magenta.  \label{p11d13}}
\end{figure}

For the first time electrocouplings of $P_{11}(1440)$ and $D_{13}(1520)$ resonances were obtained in independent analyses of N$\pi$ (0.2 GeV$^2$ $<$ Q$^2$ $<$ 5.0 GeV$^2$) \cite{Az09} and $\pi^+\pi^-p$ (0.25 $<$ Q$^2$ $<$ 0.6 GeV$^2$) \cite{Mo12} exclusive pion electroproduction  off protons. Recently these electrocouplings have been determined from the CLAS $\pi^+\pi^-p$ electroproduction data \cite{Ri03} in a range of Q$^2$ from 0.5 to 1.5 GeV$^2$. The published \cite{Az09,Mo12} and preliminary results
are shown in Fig.~\ref{p11d13}. The $P_{11}(1440)$ and $D_{13}(1520)$ electrocouplings determined from $N\pi$ and $\pi^+\pi^-p$ channels are consistent. Consistent results on electrocouplings of prominent N$^*$-states
determined in
independent analyses of major meson electroproduction channels with different backgrounds
strongly suggest the reliable extraction of these fundamental quantities. Masses, total decay widths of $P_{11}(1440)$ and $D_{13}(1520)$  resonances and
their branching fractions to the $\pi\Delta$ final states determined in our analysis \cite{Mo12} are consistent with the RPP results
\cite{rpp} and coupled-channel analysis by Bonn-Gatchina group \cite{BG12}.

Analyses of the $P_{11}(1440)$ and $D_{13}(1520)$ electrocouplings within the framework of quark models \cite{Az07,Az12,Ca95,Gia12} and a coupled-channel approach \cite{Lee13} revealed a combined contribution to the structure of these states from both an external meson-baryon cloud and an inner core of three dressed quarks in the first radial excitation and an orbital excitation of L=1 for the $P_{11}(1440)$ and $D_{13}(1520)$ states, respectively. The studies of these electrocouplings offer further insight to resonance structure allowing us to explore how the bound systems of three dressed quarks with different quantum numbers are generated and how the meson-baryon cloud depends on the isospin and spin-parity of the states. A successful description of quark core contribution to $P_{11}(1440)$ electrocouplings starting from QCD Lagrangian was achieved in the exploratory DSEQCD approach \cite{Wi12}, which well reproduces the estimates for quark core contributions obtained from the results on $P_{11}(1440)$ electrocouplings employing the coupled-channel approach \cite{Lee10}.

\begin{figure}[htp]
  \includegraphics[width=3.7 cm]{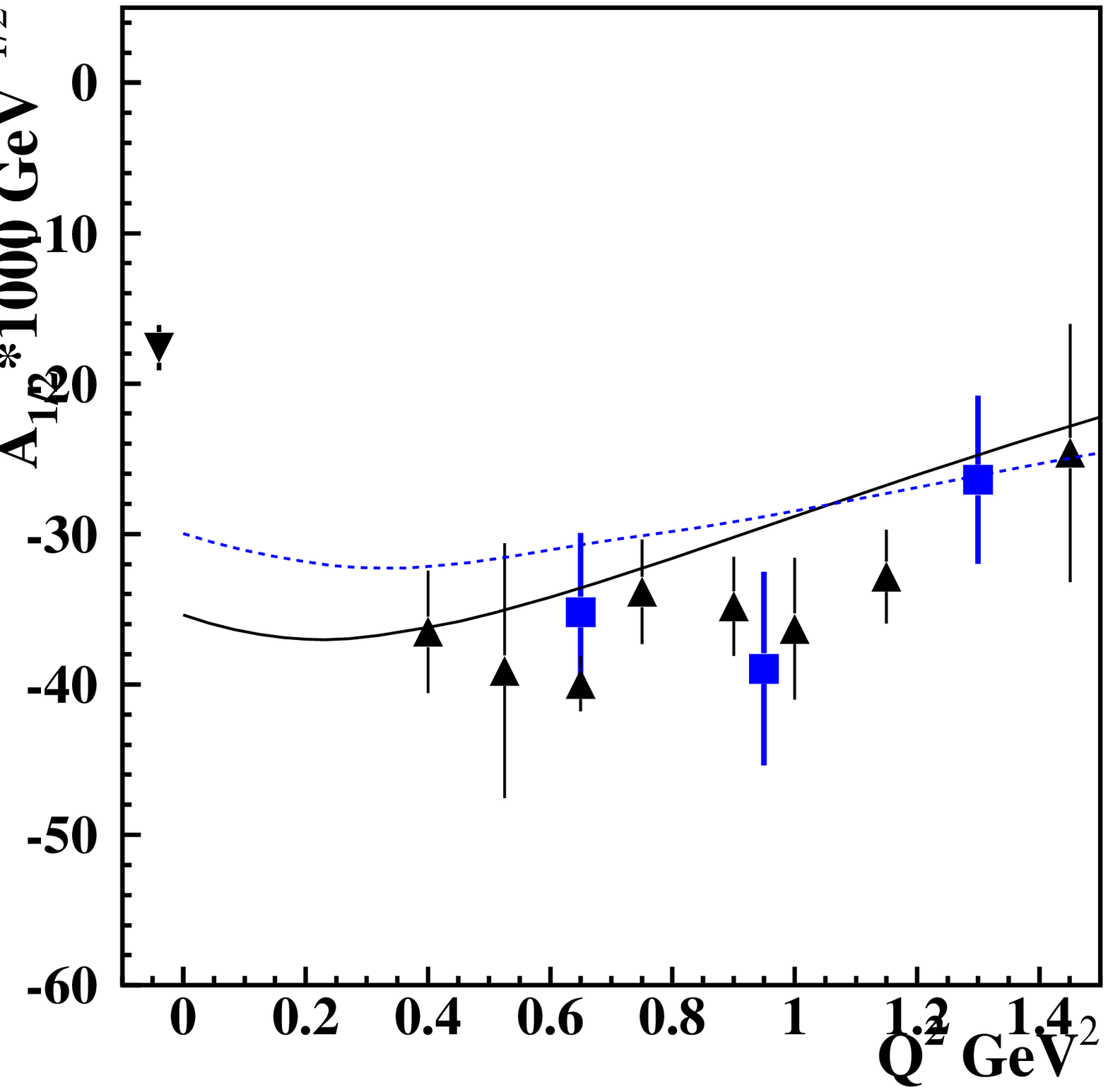}
  \includegraphics[width=3.7 cm]{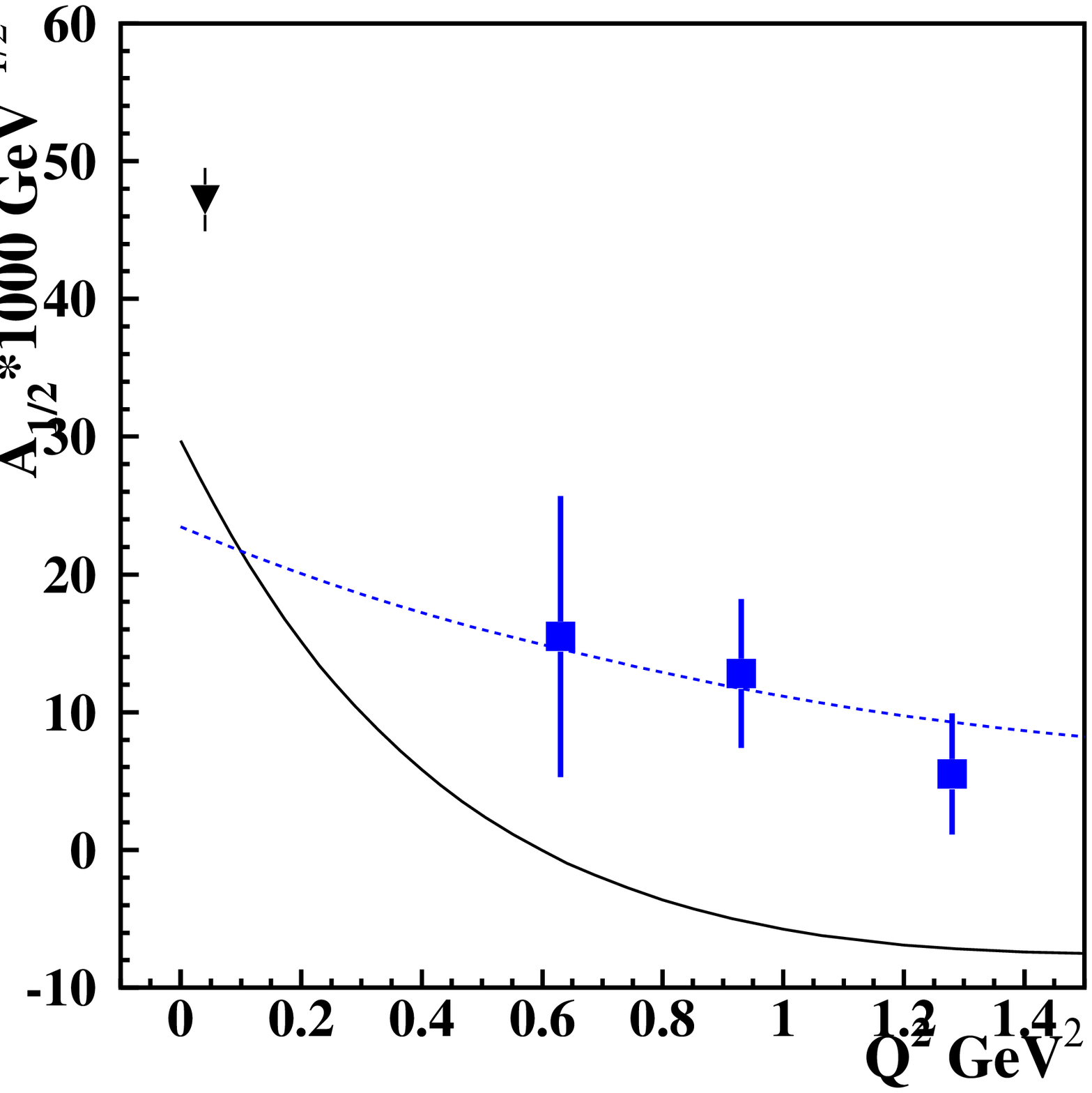}
  \includegraphics[width=3.7 cm]{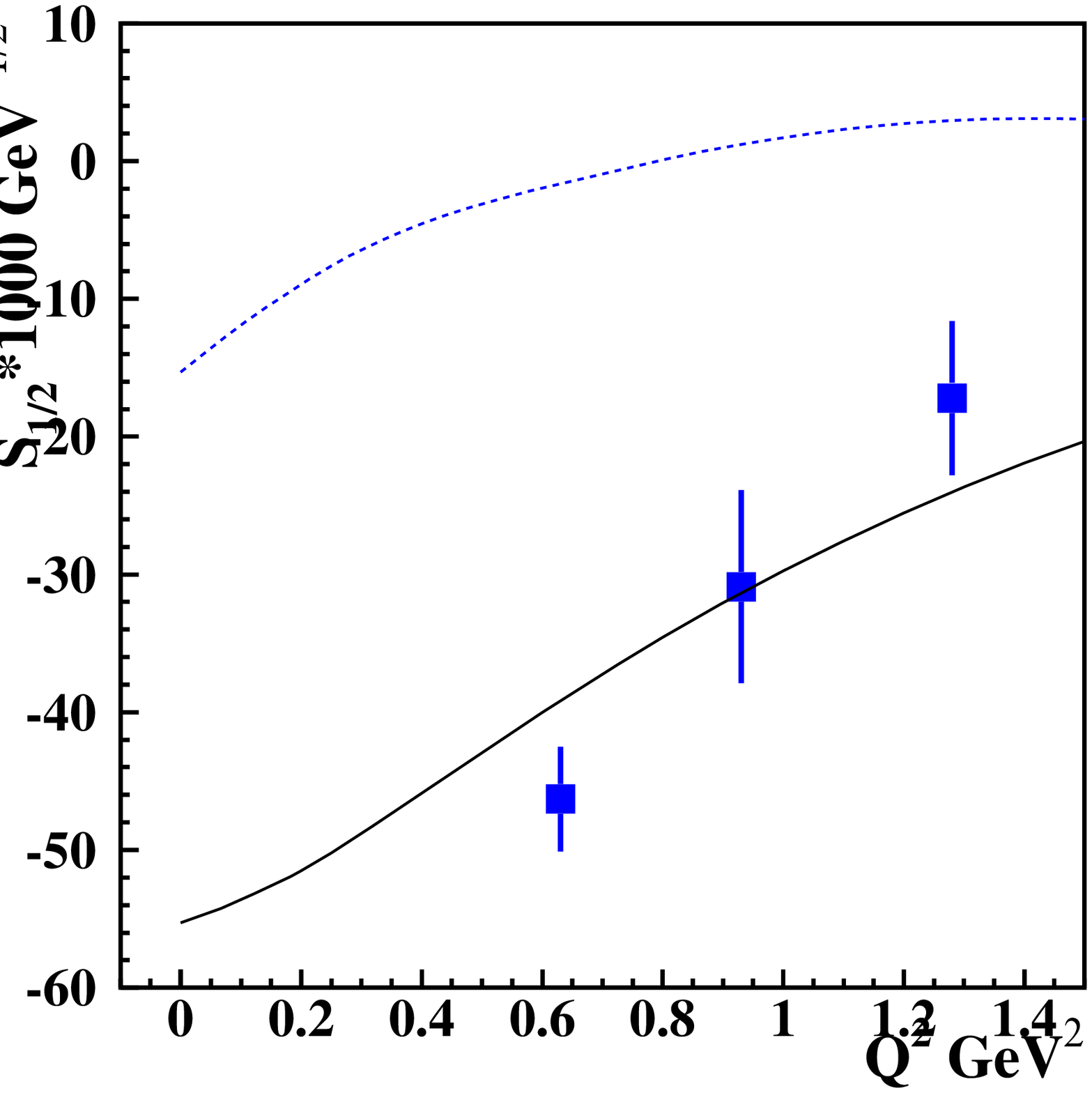}
\vspace*{8pt}
\caption{(Color online) Left: $A_{1/2}$ electrocouplings of $F_{15}(1680)$ resonance. The MAID results from analysis of N$\pi$ electroproduction data are shown as triangles \protect\cite{Tia11}, while squares stand for preliminary CLAS results from  the $\pi^+\pi^-p$ electroproduction data \protect\cite{Ri03}. Preliminary CLAS results on $A_{1/2}$ and 
$S_{1/2}$ electrocouplings of $S_{31}(1620)$ from the $\pi^+\pi^-p$ electroproduction data \protect\cite{Ri03} are shown in the middle and right panels, respectively. The evaluation of these electrocouplings within the framework of a quark model \protect\cite{Gia12} and a Bethe-Salpeter approach \protect\cite{Ro13} are shown as solid and dashed lines. \label{s31f15}}
\end{figure}

 Preliminary results on electrocouplings of $S_{31}(1620)$, $S_{11}(1650)$, $F_{15}(1685)$, $D_{33}(1700)$, and $P_{13}(1720)$ 
 states were
 obtained from the CLAS $\pi^+\pi^-p$ electroproduction data \cite{Ri03}. Most of the excited states with masses above 1.6 GeV decay preferentially to the final N$\pi\pi$ states, making the  $\pi^+\pi^-p$ electroproduction data the primary source of information on electrocouplings of high-mass resonances. Examples of available high-mass state electrocouplings are shown in Fig.~\ref{s31f15}. Analysis of  $\pi^+\pi^-p$ channel confirmed the $F_{15}(1685)$ electrocouplings determined previously from N$\pi$ electroproduction data \cite{Tia11} and provided accurate results on the transverse 
 electrocouplings and the first information on the longitudinal
 electrocouplings of all the aforementioned high-lying excited proton states.  Analyses of these electrocouplings within the framework of quark models \cite{Gia12,Ro13} demonstrated that the transition amplitudes to  N$^*$ and $\Delta^*$ states of different spin-parity and isospin further extend our knowledge on the N$^*$ structure. The models \cite{Ca95,Gia12,Ro13} have limited success in describing resonance electrocouplings. None of them is able to reproduce the electrocouplings of these resonances.

\section{Conclusion}

Electrocouplings of a large number of excited nucleon states in the mass range up to 1.8 GeV have become available from analyses of exclusive meson electroproduction off protons. Consistent results on resonance electrocouplings from independent analyses of major exclusive channels with different non-resonant contributions strongly suggest the reliable extraction of these fundamental quantities from the data. The considerable extension of the information on resonance electrocouplings provides new opportunities to explore, how N$^*$ states of different quantum numbers are generated by strong interaction in the non-perturbative regime.




\end{document}